\documentclass[11pt]{article}
\usepackage{moriond,epsfig, amsmath}

\begin{document}
\vspace*{4cm}
\title{The Chiral Magnetic Effect: \\
Measuring event-by-event $\mathcal{P}$- and
$\mathcal{CP}$-violation with heavy ion-collisions}

\author{Harmen J.\ Warringa}

\address{Institut f\"ur Theoretische Physik, Goethe Universit\"at, \\
Max-von-Laue-Str. 1, 60438, Frankfurt am Main, Germany}

\maketitle
\abstracts{Gluon field configurations with nonzero topological charge
  induce $\mathcal{P}$- and $\mathcal{CP}$-odd effects. Such
  configurations are likely to be produced during heavy ion
  collisions. In this article, I will argue that in the intense
  (electromagnetic) magnetic field produced in non-central heavy ion
  collisions, topological charge creates an electromagnetic current in
  the direction of the magnetic field. This is the Chiral Magnetic
  Effect.  It leads to separation of positive from negative charge
  along the direction of the magnetic field. I will point out that
  this effect can be investigated experimentally with a charge
  correlation study and will refer to interesting data from the STAR
  collaboration. }

\section{Introduction}
The goal of the study of heavy ion collisions is to understand the
properties of nuclear matter under extreme circumstances. The
fundamental theory that describes the relevant dynamics during such
collisions is Quantum Chromodynamics (QCD). QCD contains really a lot
of intriguing features. One of them is its connection to topology.
Another highlight is quantum mechanical symmetry breaking (also called
anomaly). And the last example I want to mention here is the existence
of parity ($\mathcal{P}$) and charge-parity ($\mathcal{CP}$)-odd
effects. In this article I will discuss these examples in more detail
and show that they are intricately linked to each other.  Since heavy
ion collisions probe QCD, these effects should somehow play a role
during the collisions. The question which then immediately pops up is:
how big is this role and how can we observe it.  I will try to answer
these two questions to some extent in this article.

\section{Topology in QCD}
The vacuum (i.e.\ state with lowest energy) of a non-Abelian field
theory like QCD has non-trivial structure \cite{CDG76}. It turns out
that there are an infinite number of different vacua that all can be
characterized by an integer winding number $n_\mathrm{W}$. This
winding number is a topological invariant, that means that a smooth
deformation of the vacuum state while staying in the ground state
cannot change $n_\mathrm{W}$.  One needs energy to change
$n_\mathrm{W}$, therefore all vacua classified according to their
winding number are separated by a potential barrier.  A gauge field
configuration with nonzero topological charge interpolates between
these distinct vacua and hence probes the potential.  The topological
charge $Q$ is given by \cite{BPST}
%\begin{equation}
$  Q = \frac{\alpha_s}{8\pi } \int \mathrm{d}^4 x \, 
  F^a_{\mu\nu} \tilde F^{a \mu \nu},$
%\end{equation}
here $F_{\mu \nu}^a$ is the gluon field-strength tensor and $\tilde
F^{a \mu \nu} = \tfrac{1}{2} \epsilon^{\mu \nu \rho \sigma} F^a_{\rho
  \sigma}$ its dual, with the complete antisymmetric tensor
$\epsilon^{0123} = 1$. In general $Q$ is not quantized, but if the
gluon field configuration starts from a vacuum at $t=-\infty$ and ends
in a vacuum at $t=\infty$, one can show that $Q$ is equal to the
change in winding number and hence an integer, i.e. $Q =
n_\mathrm{W}(t=\infty) - n_\mathrm{W}(t=\infty)$.

One possibility of interpolating between two different vacua is by
quantum tunneling through this potential barrier. The relevant
configurations in this case are called instantons \cite{BPST}, and the
tunneling rate (which is equal to the Euclidean topological
susceptibility $\chi_E = \langle Q^2 \rangle / V$) is proportional to
$\exp(-2 \vert Q\vert/ \alpha_s)$ \cite{H}. At low energies and
temperatures this rate is sizable, from the Witten-Veneziano relation
it follows that $\chi_E \approx (180\; \mathrm{MeV})^4$.  At very high
energies where perturbation theory is valid, this rate becomes very
small so that in that case instantons can safely be neglected.  The
instantons are even more suppressed at high temperatures due to
screening~\cite{GPY}. But then also a new possibility appears which is
traversing over the barrier in real-time.  The relevant configurations
are sphalerons (originally discussed for
weak-interactions~\cite{sphaleron}, but also existent in
QCD~\cite{MMS}), and the rate (real-time topological susceptibility)
is at high temperatures much less suppressed (since it does not invoke
tunneling) and proportional to $\alpha_s^5 T^4$ with a large
prefactor~\cite{BMR}. This rate means that one could expect of order
several transitions per $\mathrm{fm}^{-3}$ per $\mathrm{fm}/c$ in the
deconfined phase. In strongly coupled supersymmetric Yang-Mills theory
the sphaleron rate is sizable too, as was found by applying the
AdS/CFT correspondence \cite{SS}.  The discussed rates are for thermalized
isotropic systems. To obtain the rate of production of topological
charge in heavy ion collisions one should also take into account the
collision geometry and the fact that equilibrium might not have been
achieved. This can change these estimates, especially just after the
collision when the quark gluon plasma has not yet been
formed~\cite{KKV}.

\section{Axial anomaly}
Hadrons which are build out of quarks are the QCD states measured in
the detectors. So in order to find experimental evidence for
topological charge it is necessary to understand how topological
charge deals with quarks. While in the limit of vanishing quark masses
axial $\mathrm{U}(1)$ is a symmetry of the QCD Lagrangian, it is
broken by quantum effects. This quantum mechanical symmetry breaking
(or axial anomaly) gives rise to an exact identity.  In the limit of
zero quark mass this identity reads for each flavor separately
\cite{anomaly}
%\begin{equation}
$\Delta N_5 \equiv \Delta (N_R - N_L) = - 2 Q$
%\label{eq:axialanomaly}
%\end{equation}
where $N_{R,L}$ denotes the number of quarks minus antiquarks with
right/left-handed chirality. The quantity $\Delta N_5$ is the change
in chirality in time. A particle with right-handed chirality has
right-handed helicity, while an anti-particle with right-handed
chirality has left-handed helicity.  Right (left)-handed helicity
means spin and momentum (anti)-parallel. The difference $N_R
- N_L$ can also be read as the total number of quarks plus antiquarks
with right-handed helicity minus the total number of quarks plus
antiquarks with left-handed helicity.

So topological charge induces chirality by the axial anomaly. Here we
see the three intriguing features mentioned in the introduction
coming together. Let us dive a little deeper into the relation between
topological charge and $\mathcal{P}$- and $\mathcal{CP}$-violation.

%The reason that the axial $\mathrm{U}(1)$ symmetry is broken in the
%limit of zero quark mass, is due the anomaly, but occurs only if at
%the same time topological charge plays a role. One observes the
%breakdown of axial $\mathrm{U}(1)$ symmetry in the hadron spectrum;
%for example the $\eta'$ meson is much heavier than the other
%pseudoscalar mesons. This is clear evidence for the relevance of
%topological charge at low energies.

One could add the so-called $\theta$ term (to be precise $\theta Q$)
to the QCD action.  Such term gives rise to direct $\mathcal{P}$- and
$\mathcal{CP}$-violation.  Measurements of the electric dipole moment
of the neutron constrain $\vert \theta \vert$ to be smaller than of
order $10^{-10}$.  Since $\theta$ couples to topological charge
in the action, at nonzero $\theta$, $\mathcal{P}$ and $\mathcal{CP}$
are broken due to gluon fields with nonzero topological charge.  
In the case that $\theta = 0$ the probability to generate either a
gluon configuration with positive or negative topological charge is
equal so that $\mathcal{P}$ and $\mathcal{CP}$ are unbroken.

Instantons and sphalerons are objects with a certain size and
life-time (the size of the sphaleron is limited by the magnetic
screening length $1/{\alpha_s T}$). Therefore in the matter produced
in heavy ion collisions many of such configurations will be generated
at different points in space and time with different values of
topological charge. Since this is essentially a random processes, in
each event a net topological charge can be generated globally.
Only if one averages over many events one should find that $\langle Q
\rangle =0$ (if $\theta = 0$).  From the anomaly we then know how much
chirality is induced. The chirality averaged over many collision
events should vanish too.  But in an individual event a nonzero
chirality can be generated globally. One speaks in this case therefore
of event-by-event $\mathcal{P}$- and $\mathcal{CP}$-violation.

\section{The Chiral Magnetic Effect}

\begin{figure}[t]
\includegraphics[scale=0.43]{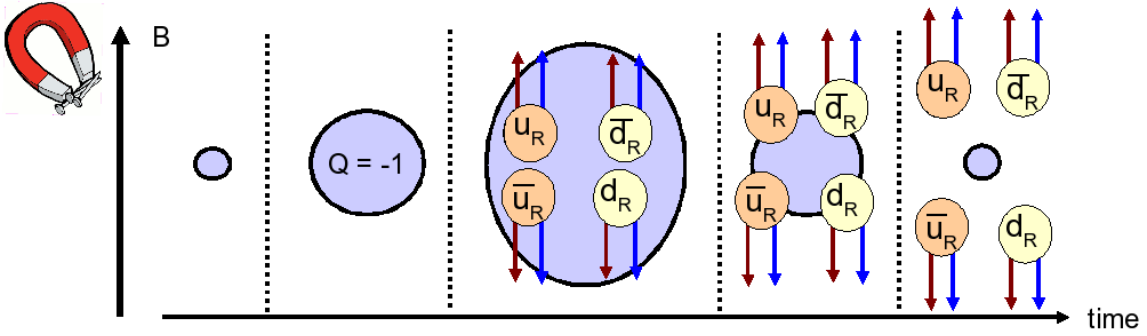}
\caption{The Chiral Magnetic effect. A gluon field with nonzero
  topological charge produces a difference between the number of
  particles plus antiparticles with right- and left-handed helicity. 
  In an external
  magnetic field  the spins (blue arrows) of these particles tend to
  align along the magnetic field and consequently the momenta (red
  arrows) align as well.  As a result an electromagnetic current is
  created in the direction of the magnetic field. 
  \label{fig:chimag} }
\end{figure}

Now that we have seen that topological charge induces chirality, let
us see how one could measure this nonzero chirality.  When two heavy
ions collide with nonzero impact parameter, a (electromagnetic)
magnetic field of enormous magnitude is created in the direction of
angular momentum of the collision (at 0.2 fm/c after the collision it
is for moderate impact parameters of order $10^3 \sim
10^4\;\mathrm{MeV}^2$ corresponding to $10^{17}\;\mathrm{G}$
\cite{KMW}).  In a background magnetic field, the quarks can gain
energy by aligning their magnetic moments along the magnetic
field. Positively charged quarks/antiquarks with right-handed helicity
have positive magnetic moment and will tend to align their spin
parallel to the magnetic field.  Since right-handed helicity means
that spin and momentum are parallel, also the momentum will be
pointing parallel to the magnetic field.  Negatively charged
quarks/antiquarks with right-handed helicity have negative magnetic
moment, and for the same reasons tend to point their momentum
anti-parallel to the magnetic field. The particles and antiparticles
with left-handed helicity will move in the opposite direction.
Therefore if a nonzero chirality is present in a background magnetic
field, an electromagnetic current will be induced in the direction of
the magnetic field. This is the so-called Chiral Magnetic effect
\cite{K06} \cite{KZ} \cite{WQM} \cite{KMW} \cite{FKW}.  I have
illustrated this effect in Fig.~\ref{fig:chimag}.

For extremely large magnetic fields so that the quarks are fully
polarized one can quickly convince oneself (use
Fig.~\ref{fig:chimag}) that the total induced current is equal to $J =
\sum_f \vert q_f \vert N_5 = - 2 \sum_f \vert q_f \vert Q$, where $f$
denotes a sum over light flavors and $q_f$ is the charge of one
particular flavor. For smaller magnetic fields, we have computed
induced current in a more general setting.  For constant and
homogeneous magnetic fields the size of the current density is
determined by the electromagnetic axial anomaly and equal to
\cite{FKW} $j = \sum_f q_f^2 \mu_5 B / (2 \pi^2)$, see also
\cite{NM}. Here $\mu_5$ denotes the chiral chemical potential, which
is used to describe a system with nonzero chirality. The chiral
chemical potential can be expressed in terms of $N_5$ by taking a
derivative of the thermodynamic potential with respect
to $\mu_5$,  $n_5 = - \partial \Omega
/ \partial \mu_5$. In this way we were able to reproduce the
large magnetic field result. For smaller magnetic fields ($B$) and large
temperatures ($T$) we obtained for the total current \cite{FKW}
(here $\mu$ denotes quark chemical potential)
\begin{equation}
 J = - \frac{3}{\pi^2}
\frac{Q}{T^2 + \mu^2/\pi^2} B \sum_f q_f^2.
\label{eq:hightempcurrentqcd}
\end{equation}

In heavy ion collisions this current leads to separation of charge
along the direction of angular momentum, which is perpendicular to the
reaction plane. This leads to nontrivial correlations between the azimuthal
angles (the angle between the particle and the reaction plane) of
charged particles. To predict the behavior of these correlations one
should compute from Eq.~\ref{eq:hightempcurrentqcd} how much charge is
separated, fold it with the rate of topological charge production and
the time-dependent magnetic field, and integrate over the reaction
volume \cite{KMW} \cite{WQM}.

An observable that measures these correlations was proposed by
Voloshin \cite{V04} and preliminary data of the STAR collaboration on
these correlations have been presented at the Quark Matter conference
\cite{IVS}. The very interesting data suggests that charge is
separated perpendicular to the reaction plane. In order to establish
observation of the Chiral Magnetic effect is important to obtain
accurate predictions on for example the impact parameter and beam energy
dependence, confront these predictions with experimental results,
and rule out other possible explanations.

\section*{Acknowledgments}
I would like to thank the organizers for the wonderful meeting and
giving me the opportunity to present this work. The ``Spirit of
Moriond'' was very stimulating.  I am grateful to Dima Kharzeev, Larry
McLerran and Kenji Fukushima for the collaboration that led to these
results.  The work of H.J.W.\ was supported by the ExtreMe Matter
Institute EMMI in the framework of the Helmholtz Alliance Program of
the Helmholtz Association (HA216/EMMI).

\end{document}